\documentclass[12pt,preprint]{aastex}

\begin{document}

\title{A Study of Compact Radio Sources in Nearby Face-on Spiral
Galaxies. I. Long Term Evolution of M83}

\author{L.A. Maddox\altaffilmark{a}, J.J. Cowan\altaffilmark{a},
R.E. Kilgard\altaffilmark{b},  C.K. Lacey\altaffilmark{c},
A.H. Prestwich\altaffilmark{b}, C.J. Stockdale\altaffilmark{d}, E. Wolfing\altaffilmark{e,f}
}


\altaffiltext{a}{The Homer L. Dodge Department of Physics and Astronomy, The University
of Oklahoma, 440 W. Brooks St., Norman, OK  73019}

\altaffiltext{b}{Harvard-Smithsonian Center for Astrophysics,
60 Garden St, Cambridge, MA, 02138}

\altaffiltext{c}{Department of Physics and Astronomy, The University of
South Carolina, 712 Main St., Columbia, SC  29208}

\altaffiltext{d}{Department of Physics, Marquette University, P.O. Box
1881, Milwaukee, WI  53201}

\altaffiltext{e}{Department of Physics and Engineering, Washington \&
Lee University, Lexington, VA  24450} 

\altaffiltext{f}{
Current Address: Directed Technologies Inc., 3601 Wilson Blvd., Suite 650, Arlington, VA
22201}

\begin{abstract}

We present analyses of deep radio observations of M83 taken with the
Very Large Array spanning fifteen years,
including never before published observations from 1990 and 1998.
 We report
on the evolution of 55 individual point sources, which include four of the six known
historical supernovae in this galaxy.  
A total of 10 sources have X-ray counterparts from a {\em Chandra} survey.
Each of these sources show  non-thermal spectral indices, and most appear to be X-ray supernova
remnants.  
Comparing the radio source list to 
surveys in optical and X-ray, we identify three optical/X-ray supernova
remnants.  
Nearly half of the detected radio sources in these observations are coincident with known {\sc Hii}
regions lying in the spiral arm structures of the galaxy.
  We also report on changes in emission from the complex nuclear region,
which has shown variability at 20cm wavelengths.  
We confirm that the peak radio emission from the nucleus
is not coincident with the known optical center.  
One lesser nuclear peak is consistent with the optical/IR nucleus.  
Previous dynamical studies of a ``dark'' nuclear
mass indicate a possible match to other radio nuclear emission regions in M83.

\end{abstract}


\section{Introduction}

M83 (NGC 5236) is a nearby SABc galaxy that is nearly face-on \citep[$i=24^{\circ}$, ][]{tal79}.
It is relatively nearby with distance estimates ranging
 from 3.75 Mpc \citep{dev79} to 8.9 Mpc \citep{san87}. 
We have opted to use the Cepheid-established distance of 4.5 Mpc \citep{thi03}.
The proximity of M83 has made it an ideal candidate for observations at all wavelengths.
The presence of large quantities of gas and dust imply observable levels of enhanced
star formation.  
Previous studies have also noted vigorous star formation in the nuclear region of the 
galaxy \citep[e.g., ][]{elm98}.
Similar star formation activity is evidenced by the discovery of six
supernovae in modern times: SN 1923A, SN 1945B, SN 1950B, SN 1957D, SN
1968L, and SN 1983N.
The results of monitoring of these historical supernovae in the radio have been presented in 
\citet{cb82,cb85,crb,eck98,eck02} and \citet{stock05}.

In addition to the radio surveys, several X-ray studies have been performed on M83.
\citet{kil02} and \citet{kil05} included it in a larger survey of nearby galaxies.  
\citet{soria02,soria03} also
presented detailed analyses of {\em Chandra} observations  of this galaxy.
Optical surveys have also provided evidence of high star formation in the
form of {\sc Hii} regions \citep{rk83} and supernova remnants
\citep[SNRs][]{blair04}.

Comparisons of compact sources at multiple wavelengths provide critical
information about late stages of stellar evolution, emission mechanisms,
the transition of supernovae to supernova remnants, etc.
In this paper we report on the results of analysis of fifteen years of radio observations of 
M83, including new, never before published data from 1990 and 1998.
  We will begin by describing the observations dating back to 1981, before the 
Very Large Array\footnote{The Very Large Array of the National Radio Astronomy Observatory is a facility of the National
Science Foundation operated in cooperative agreement by Associated Universities, Inc.}
was completed.  A brief description of the analysis techniques will then be presented in 
Section 3.
In the sections that follow, results will be presented for specific sources.  The first discussion
will be a brief report on the observed historical supernovae.  For the remaining unidentified
sources, comparisons with optical and X-ray studies will be presented to 
identify 
SNRs, {\sc Hii} regions and X-ray binaries (XRBs).
Finally, a discussion of the complex nuclear
emission will be presented.

\section{Observations and Data Reduction}

We have observed M83 with the VLA several times between 1981 and 1998.  Table \ref{obs}
summarizes each of the observations.  Reports of the 1981, 1983 and 1990 observations 
have been presented by 
\citet{cb82,cb85} and \citet{crb}.  The new 1998 observations will be presented here.
For this paper, some of the older data have been reprocessed and reimaged
in order to to allow consistent analyses and comparisons over
many years.

The first 20cm observation was performed on 10 April 1981 using 22 antennas while the
array was being moved from A to B configuration.  More 20cm observations were done on
15 and 17 December 1983 in the BnA hybrid configuration.  A 6cm observation was taken on 
14 March 1984 in the hybrid CnB configuration.  These observations were reported in \citet{cb82,cb85}
M83 was observed again at 6cm on 14 October 1990 (CnB), and at 20cm on 30 June 1990 (BnA). 
These were reported in \citet{crb}. Finally, we observed M83 in 1998 at 20cm (June 13 and 15, BnA)
and 6cm (October 31 and November 1, CnB).  \citet{crb} reported results of a 1992 20 cm 
observation taken in B configuration.  The original 1990 observation was believed to
be corrupted at the time due to the dramatic fading of SN 1957D.  In reprocessing the 
this observation, we have determined it to be good, and since the array was in the preferred
configuration, we have opted to present the observation here for the first time.

All of the data were processed using the Astronomical Image Processing System (AIPS) provided
by NRAO.  Flux calibration was performed using 3C286 as the primary source calibrator.
To correct for atmospheric phase variations, a secondary calibrator, B1354-132 was used for the
1983 and 1984 observations, and J1316-336 was used for the
1990 and 1998 data sets.  
Due to changes in the AIPS system, we were unable to reproduce adequate maps for the earliest
observations.  We have opted to use the original maps presented in \citet{cb85}.  
The remaining data were then imaged using the AIPS task IMAGR using a Briggs robustness parameter of 0.
This value has the advantage of minimizing noise while allowing for excellent point source 
detection in resultant images.  During the imaging process, we also employed self-calibration
loops using the task CALIB to improve the noise levels on the maps.
The data sets were deconvolved using Gaussian restoring beams,
the sizes of which are indicated in Table \ref{obs}.  
The deconvolved  beams were
determined using the values for the ``dirty'' beam, calculated from the Fourier transform
of the {\em u-v} plane coverage.  After imaging, the task PBCOR was run on each map to
correct for the shape of the primary beam.

\section{Data Analysis}

An initial source list was obtained using the AIPS task SAD, which searches for points in
a radio image that are higher than a specified level.  A gaussian fit is the applied to each
detected area of emission, and fits that fail are rejected by the algorithm.  
Many extended and slightly extended 
sources were listed as multiple sources.  The list was compressed to account for this.
Nuclear emission was removed from the source list.  Finally, a visual inspection of the maps
was performed in order to find sources that were rejected by the detection algorithm.  The final
list of sources along with peak flux densities at all epochs is presented in Table \ref{flux}.
Table \ref{spix} lists the spectral indices ($S\propto\nu^{+\alpha}$) of each source.

\begin{figure}
\includegraphics[width=\textwidth]{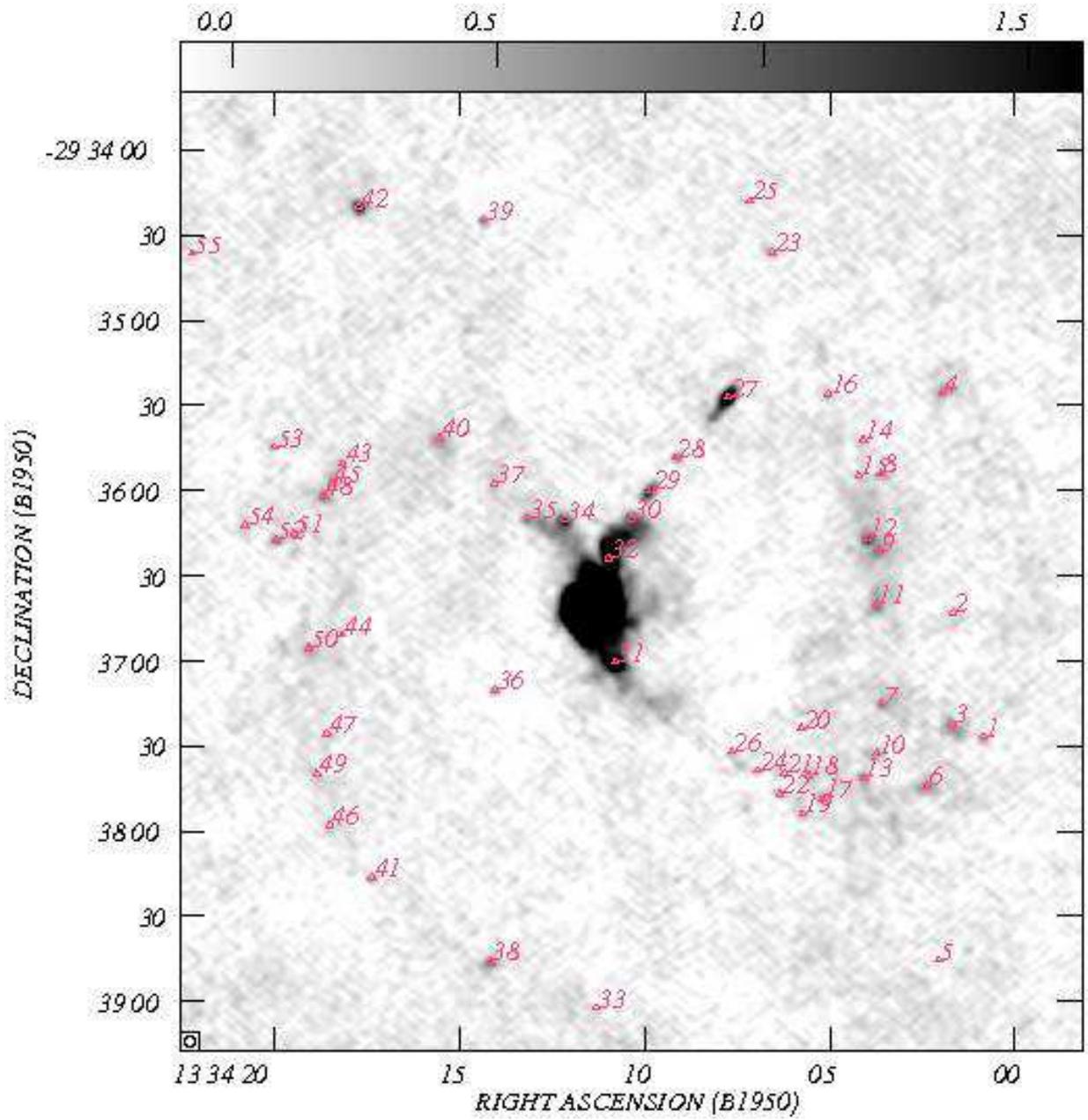}
\caption[{\sc Sources in M83}]{\label{m83source}
A 20cm map of M83 from the 1998 observation.  Sources indicated correspond to
those in  Table \ref{flux}. Greyscale spans from $-0.1$ mJy to 1.6 mJy.}
\end{figure}

\begin{figure}
\includegraphics[width=\textwidth]{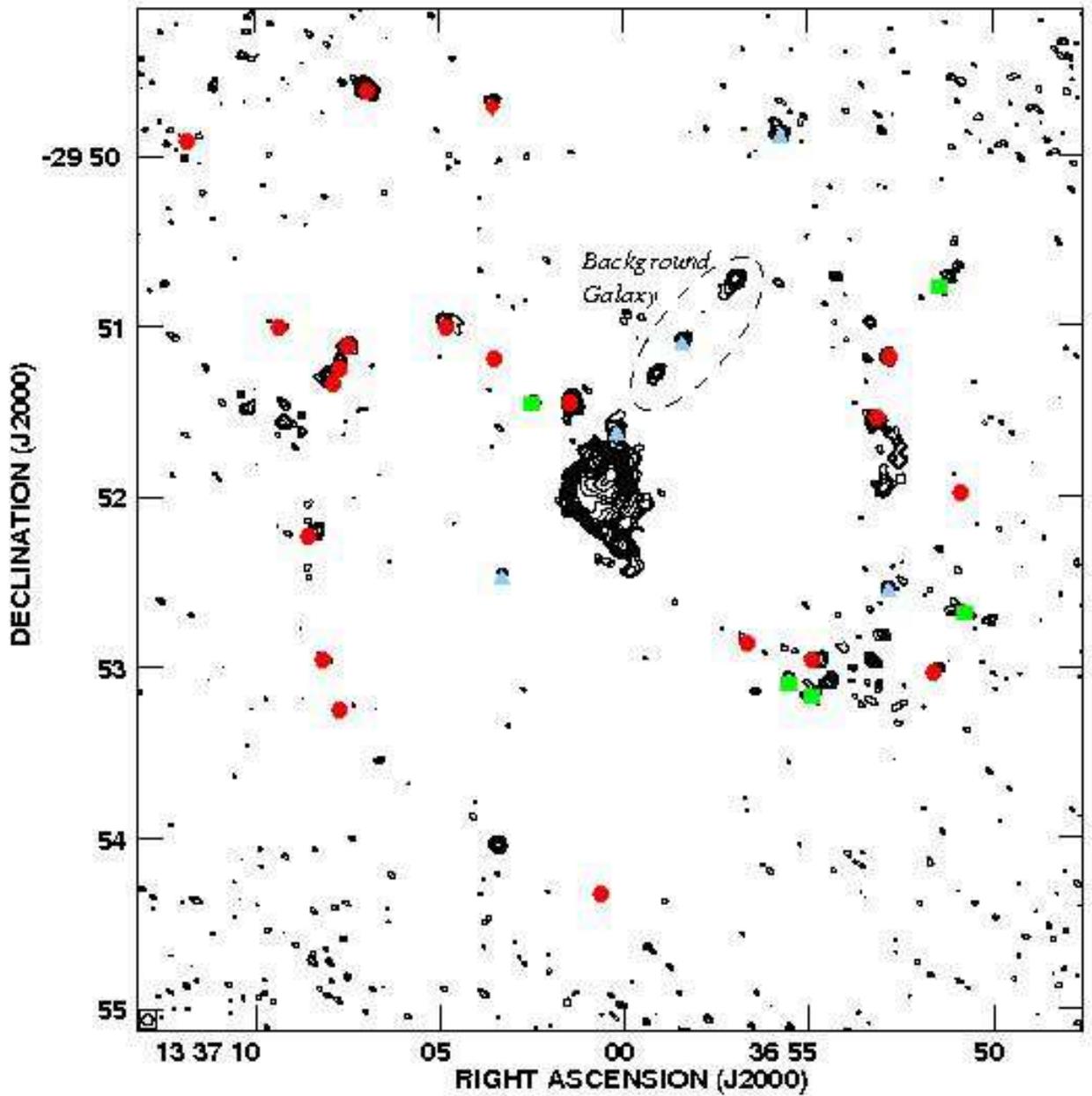}
\caption[{\sc Contours of M83}]{\label{galcont}
6cm contours of M83. Red symbols indicate radio sources with optical counterparts. Red
circles are {\scshape Hii} regions, while the red diamond indicates an optical SNR. 
Blue triangles indicate the presence of X-ray counterparts.  Green squares denote sources
that have emission in radio, optical and X-ray.
Contour levels are 0.14, 0.20, 0.28, 0.40, 0.57, 
0.80, 1.13, 1.60, 2.26, 3.20, 4.53, 6.40, 9.05 and 12.80 mJy/beam.
}
\end{figure}

The sources were then fit using the AIPS task IMFIT.  The input model was a two-dimensional 
Gaussian with a linear sloping background as was implemented in \citet{crb}.  Fluxes for the 
sources are listed in Table \ref{flux}.  Where no value exists, we assume a 3-$\sigma$ upper
limit of 0.59 mJy beam$^{-1}$ (1983 20cm), 0.14 mJy beam$^{-1}$ (1984 6cm), 0.22 mJy beam$^{-1}$ (1990 20cm),
0.11 mJy beam$^{-1}$ (1990 6cm), 0.24 mJy beam$^{-1}$ (1998 20cm), and 0.21 mJy beam$^{-1}$ (1998 6cm).  
The 20cm map shown in Figure \ref{m83source} shows the positions
of the sources in the final list.  The 6cm contour map in Figure \ref{galcont}  indicates the positions of
some of the sources discussed in \citet{crb} and \citet{eck98}.

Many of the fluxes of the sources listed in \citet{crb} differ from fluxes listed in 
Table \ref{flux}. The reprocessing and reimaging of the older data resulted in maps with 
different deconvolution sizes than those in the older maps.  This produced different flux 
measurements than maps using the older imaging algorithms.
The advantage
in reprocessing lies in the fact that all of the older data have been treated 
in the same manner as
the new data.

\section{Discussion}

\subsection{Radio Point Sources}

\subsubsection{Historical Supernovae}

We detected emission from four of the six historical supernovae (SNe) in M83 (SNe 1923A,
1950B, 1957D and 1983N).  Source designations for each of the SNe are provided in Table
\ref{flux}.  A more detailed discussion of the nature of the historical SNe is presented
in \citet{stock05}.

SN 1983N is detected only in the $1983-1984$ observations \citep{cb85,crb}, indicating
that it faded very quickly after maximum light.  This is consistent with the identification of
 SN 1983N as a Type Ib supernova \citep{weil86,boff99}.

\citet{eck98} reported the discovery of SN 1923A in the radio.
The 1990 observations showed a
very faint radio point source, and the reprocessing of the $1983-1984$ data did hint at the
presence of a source at the position of the supernova, though the flux level is too low
to call it a detection.  In the 1998 data, we again see
a faint source that is coincident with the supernova.  
In the 1990 maps, the emission appears to be non-thermal in nature. 
 \citet{rk83} report the presence of an optical 
{\scshape Hii} region at the position of source 59, coincident with the reported position
of SN 1923A.  \citet{boff99} could find no optical emission consistent with a supernova
expansion shock in the area around SN 1923A.  It is possible that we are detecting emission 
from both objects, though the radio emission is near the limits of detection for the
observations.

\begin{figure}
\begin{center}
\includegraphics[width=5in]{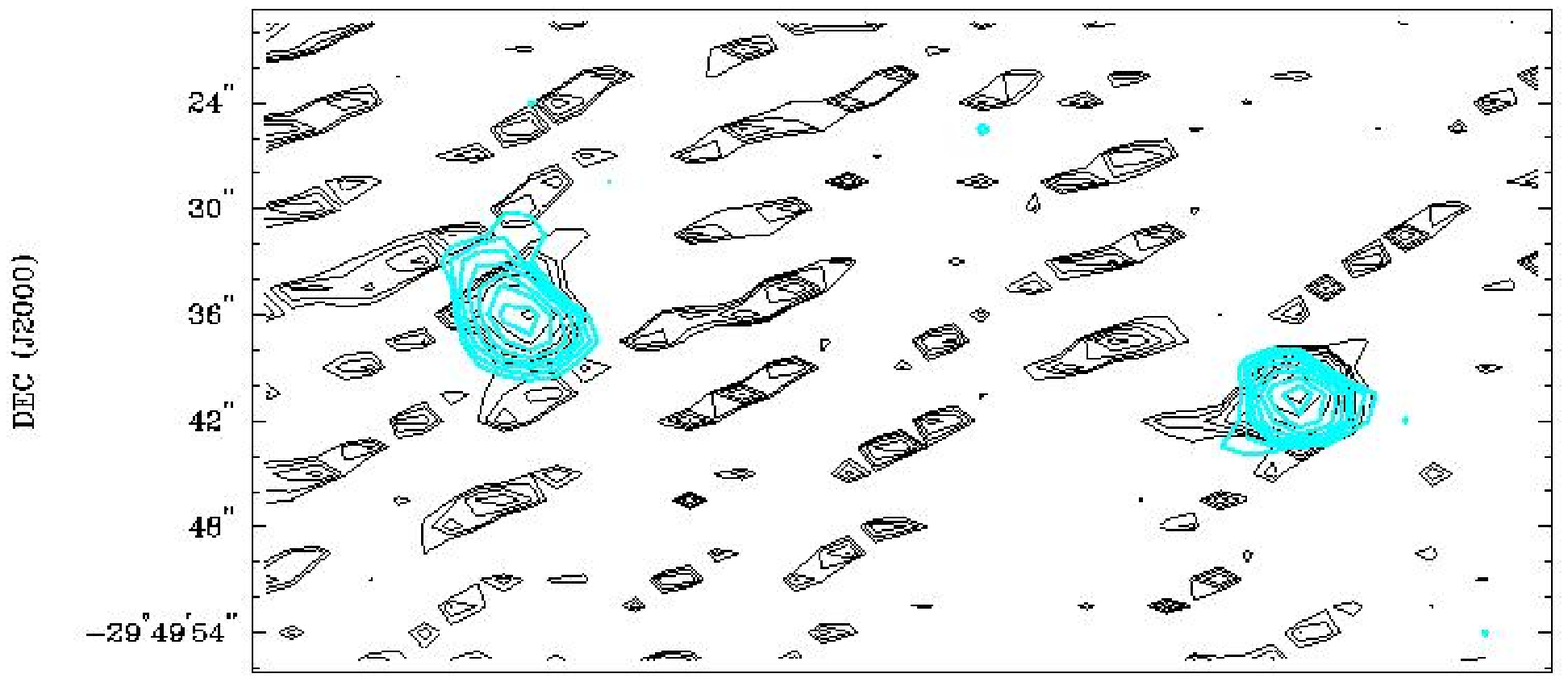}\\
\includegraphics[width=5in]{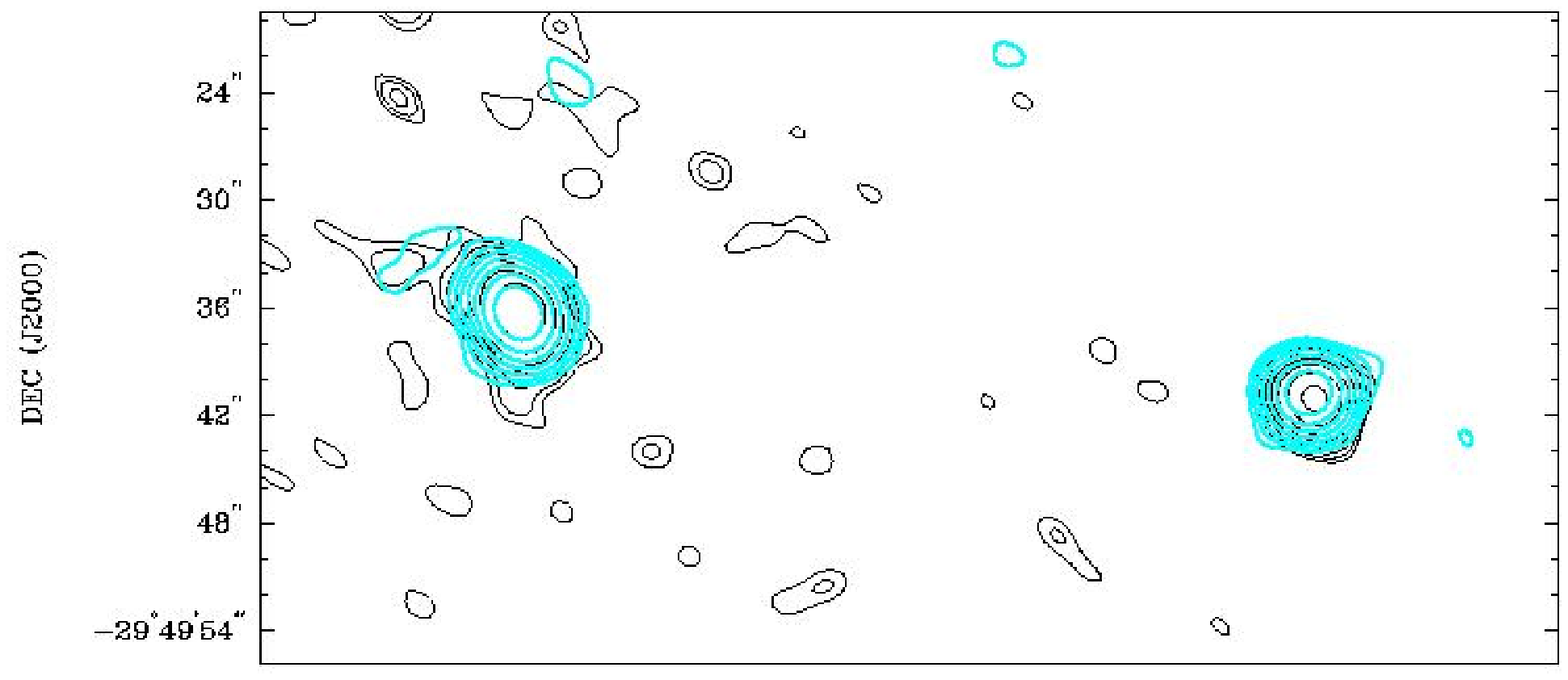}\\
\includegraphics[width=5in]{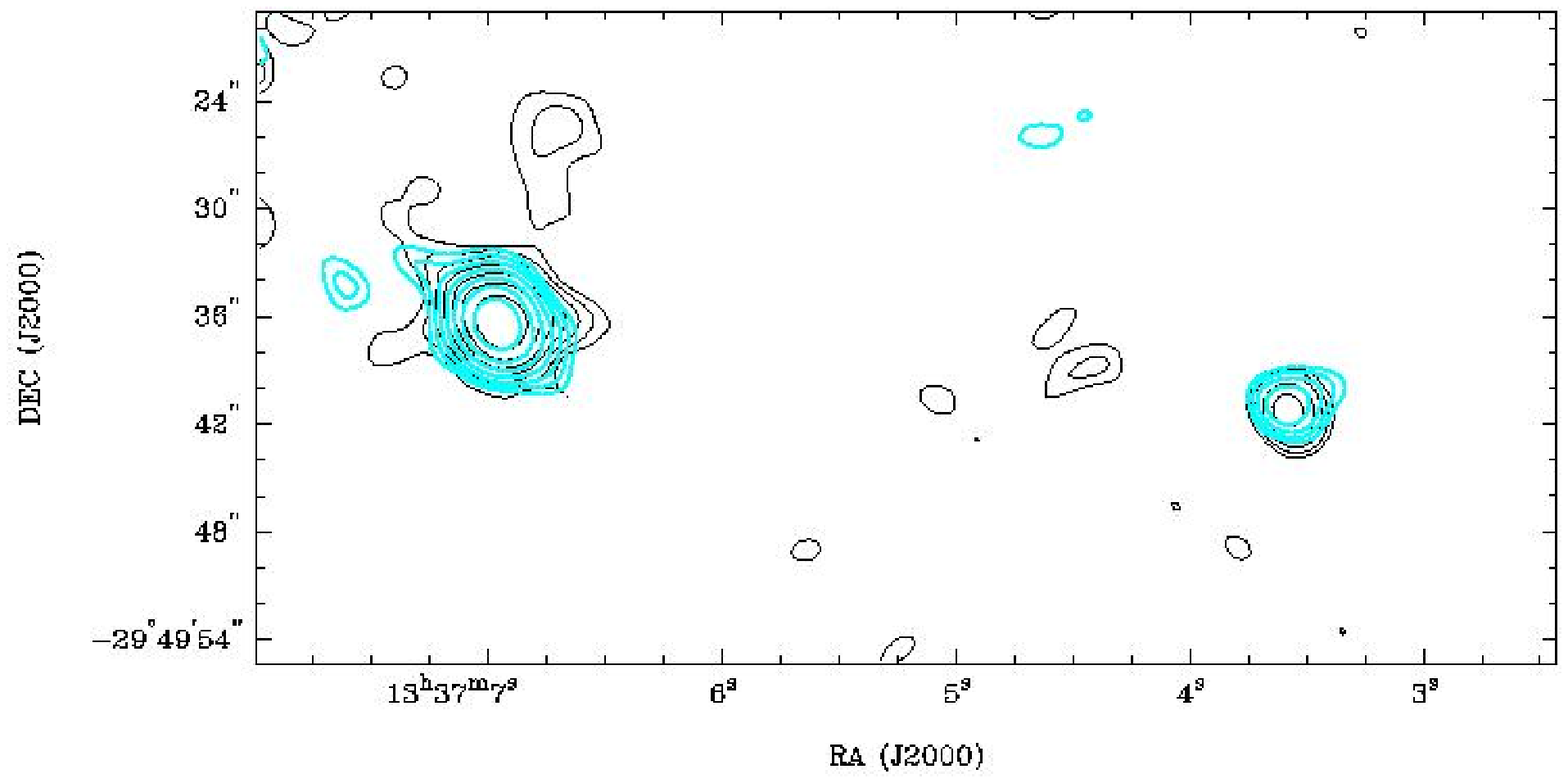}
\end{center}
\caption[{\sc Time Evolution of SN 1957D}]{\label{57dtime}
A series of contour plots showing the time evolution of SN 1957D. Black contours are 20cm
emission and thick blue green contours are 6cm emission.
Contour levels for both plots are 0.20, 0.28, 0.40, 0.57, 0.8, 1.13, 1.60 and 2.26 mJy/beam.
SN 1957D is the right hand source.  The left hand source is an {\scshape Hii} region. Time
runs from top to bottom.
}
\end{figure}

\cite{crb} and \citet{stock05} noted a continuing decline in the radio emission from 
SN 1957D.  However 
there has been unusual optical activity at the position of the explosion. \citet{long89}
reported the detection of an optical remnant of SN1957D.  The remnant  then faded below
detectable limits within a few years \citep{long92}, suggesting a possible drop in the
circumstellar density through which the explosion shock is propagating.  Our radio observations
also indicate the rapid decline of SN 1957D (see Figure \ref{57dtime}), and the spectrum 
seems to be flattening.
This could indicate that the supernova is approaching the level of the {\scshape Hii}
region in the area.  The supernova has not faded below this level, but the flattening
of the spectrum could mean a greater contribution from the thermal emission of the gas.

\begin{figure}
\begin{center}
\includegraphics[width=3in]{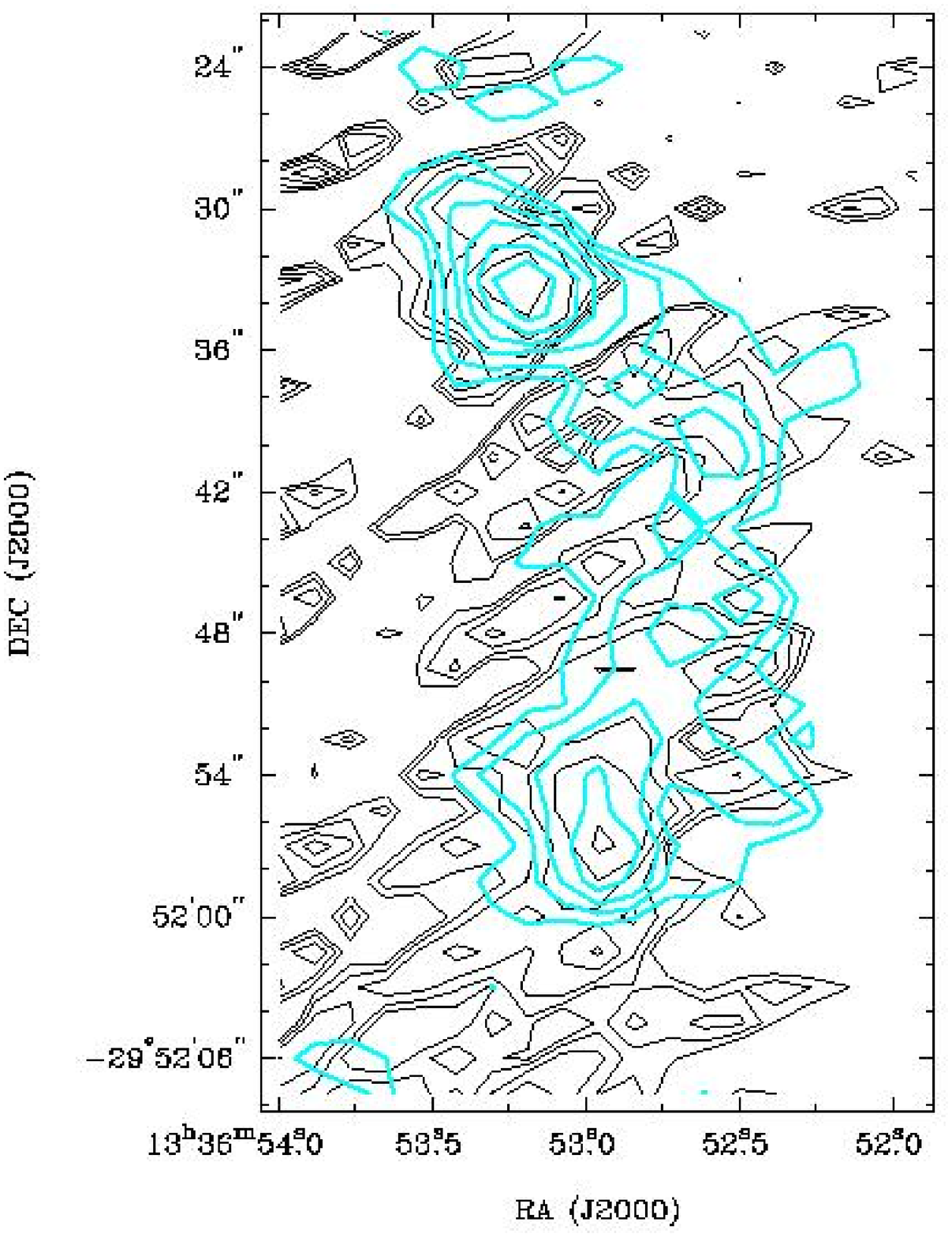}
\includegraphics[width=3in]{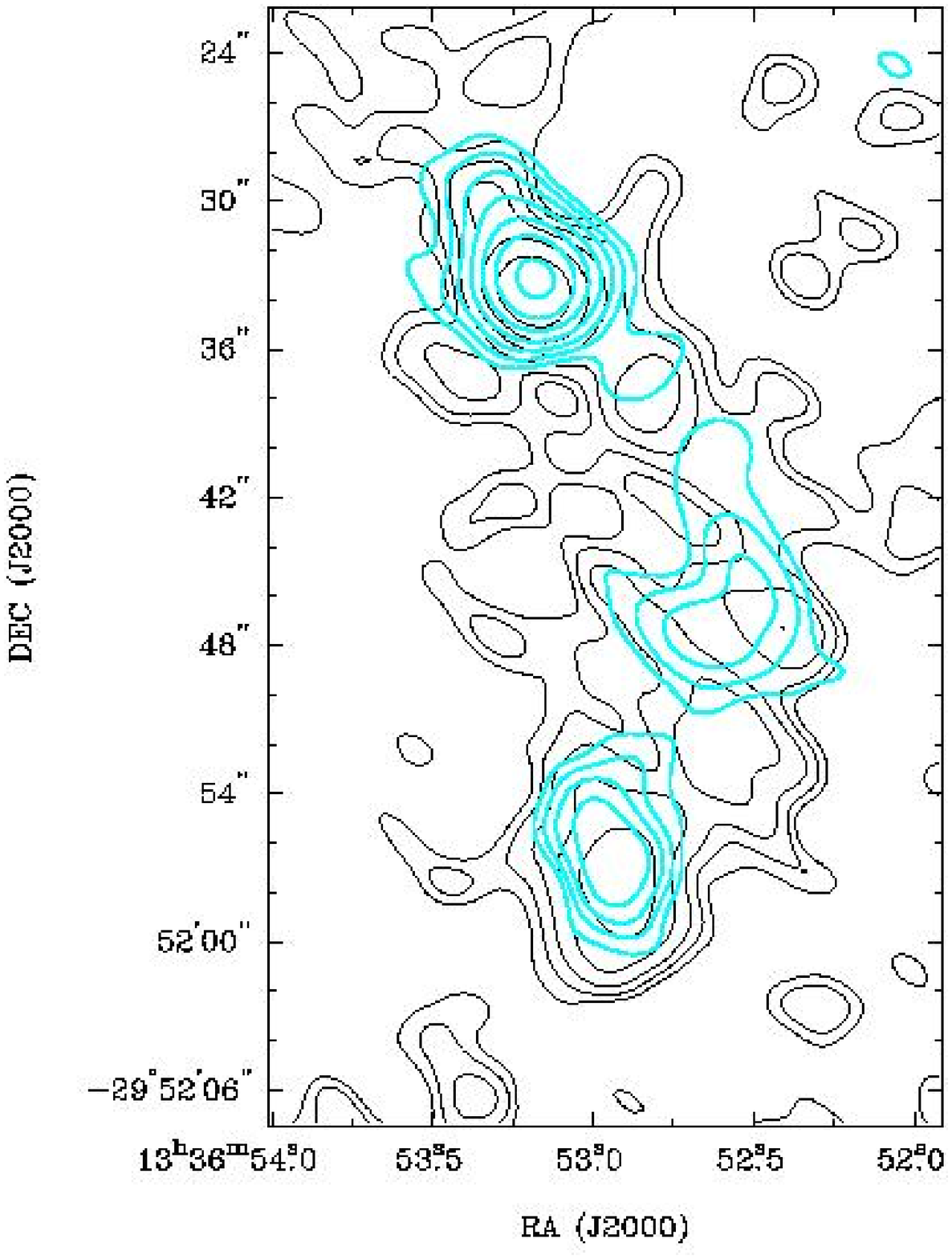}
\includegraphics[width=3in]{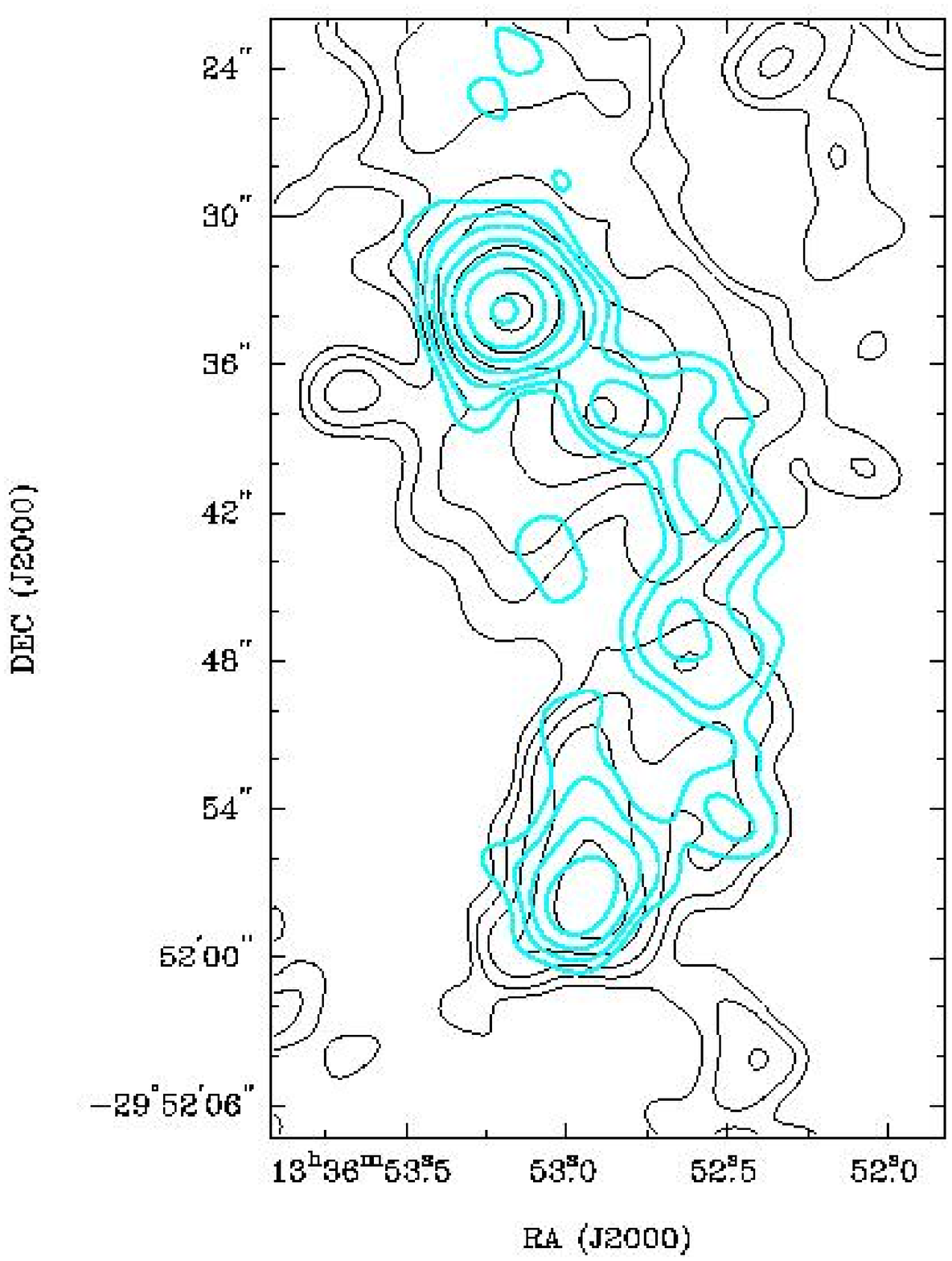}
\end{center}
\caption[{\sc Time Evolution of SN 1950B}]{\label{50btime}
Same as Figure \ref{57dtime}, but near the region of SN 1950B.  Contour levels are the same.
SN 1950B is the lower of the two major sources. Time progress clockwise form the upper left.
}
\end{figure}

As reported in \citet{stock05}, the radio spectrum of SN 1950B seems to have
flattened and  the flux has remained relatively constant across the 1990 to 1998 epochs.  This would imply that the 
supernova has faded below the level of the local {\scshape Hii} regions. SN 1950B occurred
in a complex region of emission.  Figure \ref{50btime} shows the region and how the emission
has changed over the course of the observations.

\subsubsection{Supernova Remnants}

After studying the historical supernovae, we turned our attention to searching for radio
counterparts to the numerous optical SNRs reported by \citet{blair04}.
In our search we identified four sources coincident with optical SNRs.  Three of these
radio sources had corresponding X-ray sources as reported in \citet{soria02,soria03}.
The SNR counterparts are listed in Table \ref{snrcount} and the positions are indicated
in Figure \ref{m83source}.  With the exception of source 39, all
sources exhibit continuum spectral indices consistent with those due to synchrotron
emission models for radio emitting SNRs \citep{bere04}.

Source 19 is located on the edge of the southern spiral arm of M83.  It has exhibited inverted
spectra throughout the observations, with some flattening in 1998. The source faded at 6 cm
from 1983 to 1990, and it has remained constant in flux between 1990 and 1998.
It could be again that we are seeing larger contributions from thermal emission in this region.
The large uncertainties in the spectral index measurements stem from the faintness
of the source.  \citet{soria03} note the detection of a faint X-ray source coincident
with source 19.  
Two X-ray colors are obtained: a hard color (HC),  
defined as $(H-M)/(H+M)$ and a soft color (SC), defined as $(M-S)/(M 
+S)$; where S, M and H are counts in the soft (0.3-1.0 keV), medium  
(1.0-2.0 keV) and hard (2.0-8.0 keV) bands, respectively.  
Our resulting colors ($HC=-0.19\pm0.45,\ SC=-0.68\pm0.17$) are consistent with SNR emission
using the classification scheme in \citet{prest03}.
Our radio measurements appear to confirm the identification of this
source as an SNR.

Source 22 has shown a consistent non-thermal spectrum throughout all observational
epochs.  It lies in the same region of the galaxy as source 19.  This is a crowded
region of emission along the southern spiral arm, making isolated measurements difficult.
Source 22 has steepened due to the greater fading in the 6cm maps.  The
source has shown fairly steady emission in the 20cm band. Like source 19 there is a
coincident X-ray source. Both source 19 and 22 also exhibit high H$\alpha$ luminosities
\citep{blair04}.  This is consistent with there being a denser interstellar medium
in the region of the sources.  The lower resolution 20 cm map presented in \citet{crb} indicates
a high  {\scshape Hi} density in the area of sources 19 and 22. The X-ray colors for source 22 are not typical for
an SNR ($HC=0.15\pm0.22$, $SC=-0.51\pm0.14$), though they might indicate the presence of and
XRB or an X-ray pulsar.  

Source 35 has exhibited fairly steady fluxes through all epochs that we are considering.
We have determined spectral indices for this source that are consistently non-thermal in 
nature. Our {\em Chandra} observations \citep{kil05} show that this source has a very soft spectrum
($HC=-0.82\pm0.22$, $SC=-0.53\pm0.11$), indicative of SNR emission.

Source 39 differs from the other sources discussed here in that it has exhibited
consistently thermal spectral indices over the time of observation.  We note that the flux
density of the source has remained fairly steady with some slight fading in the 6cm band.
This is more consistent with the behavior of an {\sc Hii} region than a SNR.
There is also a lack of X-ray emission from this source. \citet{blair04} identified
the optical source as an SNR candidate based on enhanced [{\sc Sii}]:H$\alpha$ ratio.
Further spectroscopic confirmation was not possible due to the relatively low luminosity 
of the region.

\subsubsection{H II Regions}

It should come as no surprise that the largest number of radio sources in M83 should be
associated with {\scshape Hii} regions.  M83 is known for exhibiting vigorous star formation
activity as indicated by the large number of observed SNe and SNRs.
By far the largest number of optical counterparts to our radio sources do
consist of
{\scshape Hii} regions.  Nearly half of the reported radio sources have coincident
optical sources identified with {\scshape Hii} emission \citep{rk83}. The radio/{\scshape Hii}
counterparts are listed in Table \ref{hiicount}.
It should be noted that the earlier observations represent some of the first detections
of extragalactic {\sc Hii} in the radio at the resolution and flux level that were
detected in M83.  This is due in part to the depth of each observation. The deep observations
also contribute to the ability to see the spiral arm structure of the galaxy as seen in Figures
\ref{m83source} and \ref{galcont}.  It is in the spiral arms where {\sc Hii} regions would
be more abundant.

It is clear from Table \ref{hiicount}, that we are preferentially identifying large
optical {\scshape H ii} regions.
\citet{rk83} establish a threshold excitation parameter $U\ge300 \mathrm{pc}\
\mathrm{cm}^{-2}$
to classify
a region as large.  Only $11\%$ of the \citet{rk83} sources are
classified as large by
this criterion, while 12 of 21 radio detected {\scshape H ii} regions fall in this
category.  If
the cut-off was relaxed to 250 pc cm$^{-2}$, $19\%$ of the \citet{rk83}
sources
would be included, while 16 of our 21 radio detected {\scshape H ii} regions now fall
in this
regime. We have restricted our comparisons to optical studies with
comparable resolutions to our
radio observations.

Most of the radio/optical counterpart sources show consistently stable flux densities and
flat spectra indicative of thermal emission processes.  This type of behavior is what would
be expected from the hot gas that composes an {\sc Hii} region.  
There are notable exceptions.
Source 27, although coincident with an optical {\scshape Hii} region, is a radio lobe
to a background FR II radio galaxy \citep{poster2,soria03}.

The spectrum of source 4 has steepened with time.  It has shown a consistent increase in 20cm
emission, while steadily fading at 6cm. In addition to the associated {\sc Hii} region, there 
exists an X-ray counterpart \citep{colb04}.  A discussion of the X-ray properties of source 4
are given in the next section.

Sources 53 and 55 are very faint sources.  This leads to large uncertainties in the spectral
index measurements for each of the sources.  As noted earlier, source 53 is coincident with the
reported optical position of SN 1923A \citep{eck98}.  The measured flux of this source is 
dependent on the noise level at that position on the map.  Similarly, the faint source 55
is greatly dependent on the rms map level at the position. While the emission is likely to 
be thermal emission from {\sc H ii}, we are unable to confidently identify source 55.

\subsubsection{\label{xray} X-ray Counterparts}

A comparison of our radio source list with the X-ray lists of \citet 
{soria03} and \citet{kil05} yield 10 X-ray counterparts. These  
sources, along with the {\em Chandra} source  designation, are listed in
Table \ref{xrcount}.  Sources 19, 22 and 35 were discussed  
previously.

\begin{figure}
\begin{center}
\includegraphics[width=4.5in]{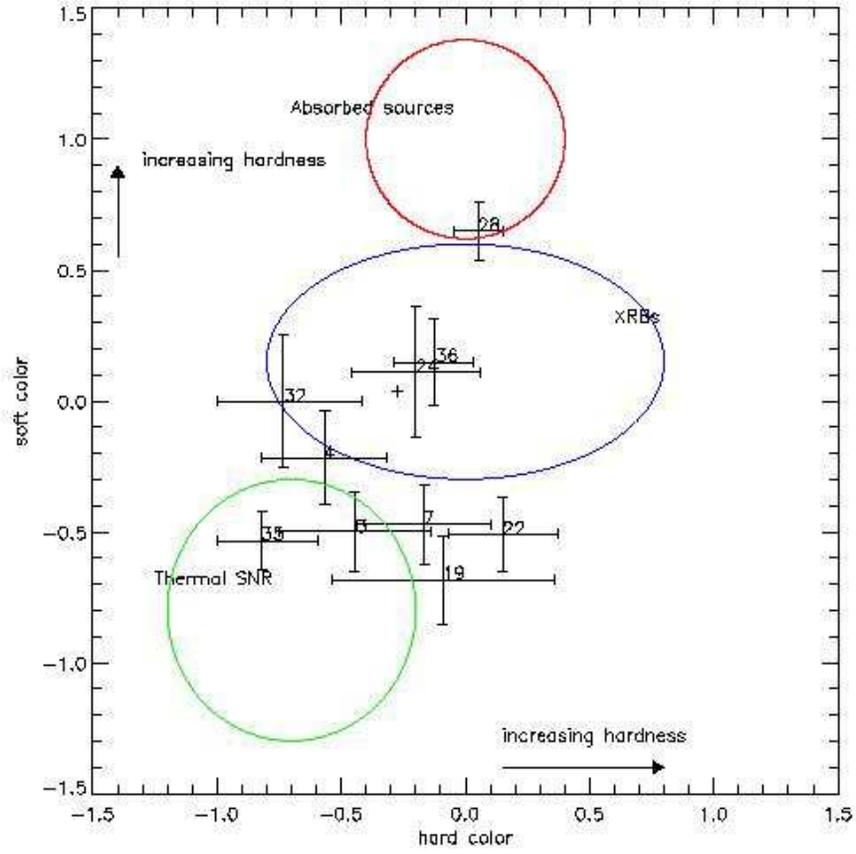}
\end{center}
\caption[{\sc X-ray color-color diagram for M83}]{\label{xrcc}
X-ray color-color diagram for the X-ray/Radio counterpart sources in M83.}
\end{figure}

In order to constrain the nature of the X-ray/radio sources, we  
examine the X-ray colors using the values published in \citet{kil05}. The X-ray  
color-color plot is presented in Figure \ref{xrcc}.  
As can be seen, a clear dichotomy exists between X-ray binary
candidates and SNR candidates:
sources 3, 4, 7, 19, 22 and 35 are SNR candidates; sources 24, 32 and
36 are XRB candidates; and
source 28 is a highly absorbed background source.
Individual sources  
are discussed below.  This same dichotomy between SNRs and XRBs was  
seen by Kilgard et al. (2006, in preparation) in M51 X-ray sources  
that are coincident with compact H$\alpha$ sources.  In M51, many of  
the SNR-candidates are also coincident with compact radio sources.   
  Thus, these observations  
help place constraints on X-ray source classification based upon X- 
ray colors.

Sources 3 and 4 show consistently non-thermal radio emission over the  
three epochs, although
for source 3 we cannot calculate a spectral index due to lack of  
detections for at least
one band in each epoch. In the radio source 3 has shown a rise in  
20cm flux density from 1983 to
1990.  That emission dropped to below detectability in 1998.  At the  
same time the source brightened
to a detectable level in 1998 after no detections in the two previous  
epochs. We do note
that there are optical {\scshape Hii} regions \citep{rk83} coincident with both   
radio sources.  Source 3 has HC$=-0.44\pm0.31$ and SC 
$=-0.50\pm0.15$, placing it in the range of thermal SNRs; source 4  
falls on the cusp between SNRs and soft X-ray binaries (likely  
LMXBs), with HC$=-0.57\pm0.25$ and SC$=-0.22\pm0.18$.  Both sources  
exhibit no variability between the two {\em Chandra} observations.

Similarly, source 7 has X-ray colors of HC$=-0.17\pm0.27$ and SC 
$=-0.47\pm0.15$, placing it also on the cusp of LMXBs and SNRs.   
Consistent non-thermal radio emission over this time strengthens the  
determination of this source as an SNR. There was an increase in 20cm  
emission, while at 6cm the flux remained flat.  There is no  
identified optical counterpart to this source so we cannot place  
other constraints on the source type.

Source 24 has X-ray colors indicative of an X-ray binary, with HC 
$=-0.20\pm0.26$ and SC$=0.11\pm0.25$.  However, the location of the  
source within a dust lane could lead to absorption of the soft X- 
rays.  As such, we cannot rule out an identification of an SNR based  
upon the X-ray data.

Source 28 has been of particular interest since it was associated  
with sources 27 and 29 \citep{crb}.
It was postulated that the sources could be part of a jet from the  
nucleus of M83.  High resolution
{\em Chandra} observations show an X-ray source at the position of  
source 28. X-ray spectral analysis
indicate that it is likely a background galaxy \citep{poster1,soria03}.   
Radio observations show
a source that  varies slightly at 20cm while fading rapidly at 6cm.
Sources 27 and 29 show consistent non-thermal emission.  At 20cm source 27 has brightened slightly, while
source 29 has remained steady or slightly faded (there is a large overlap in the uncertainty).  Both sources faded between 1990 and 1998 at 6cm.
The association of sources 27 and 29 indicate
that they are the radio lobes of a FR II radio galaxy that lies along  
the line of sight
\citep{poster1,poster2,soria02,soria03}.

Source 32 has the faintest X-ray detected counterpart to a radio  
source in M83 (L$_X < 10^{37} erg\ s^{-1}$).  The X-ray colors  
indicate that the source is a soft X-ray binary (HC$=-0.73\pm0.32$, SC 
$=0.00\pm0.26$).  The soft X-ray color may be artificially  
``softened" due to excess soft, diffuse X-ray emission near the  
source.  There is also the suggestion of X-ray variability, with the  
source not detected at all in the second {\em Chandra} observation.   
However, given the short observation time (20\% that of the longer  
observation), the source would be at or slightly below the detection  
limit. This source lies near the confused nuclear region of  the galaxy.
This leads to a large background in that area.  The measured spectral index
for source 32 is quite steep ($-1.7$), but this could be attributed to
the uncertainty in the background level.  The source could also be
exhibiting radio variability similar to the suggested X-ray variability.

Source 36 has been a mystery in previous studies.  It has no optical  
counterpart.  It is
not located in the high surface brightness  area of a spiral arm.   
The presence of the X-ray counterpart
was noted first by \citet{poster1}, and \citet{poster2} suggested  
that it was another
background galaxy.  Radio measurements show a slight rise in 20cm  
emission from the earliest epoch.
   Over the same span of time the 6cm has dramatically dropped.   
Further analysis of the X-ray indicated that  the source is  
consistent with an X-ray binary (HC$=-0.13\pm0.16$, SC$=0.15\pm0.16 
$), with evidence for X-ray spectral evolution between the two  
observations without a change in flux.

As illustrated in Figure \ref{m83source} and listed in Tables \ref{flux}, \ref{snrcount},
\ref{hiicount}, and
\ref{xrcount}, we detect
radio
emission from 55 point source sources overlapping with 6 X-ray only
sources,  20
optical only sources, and 4 X-ray/optical sources.  $54\%$ of the sources
in
our radio sample
have counterparts in the optical and X-ray, which can be compared to the
SNR populations
measured by \citet{pan02} for M33, NGC~300, and NGC~7793.  For M33, they
report radio
emission from 53 SNRs with 37 optical and 11 optical/X-ray co-detections;
for NGC~300,
they report radio emission from 17 SNRs with one optical, one X-ray, and
two optical/X-ray
co-detections; and for NGC~7793, they report seven radio SNRs with one
each optical,
X-ray, and optical/X-ray co-detection.  It should be noted that the
resolution for
the observations was $6\arcsec$, that the statistical errors of the SNR
spectral
index measurements were so large that the spectral indices for most
identified
radio sources are not clearly identified as SNRs by their own
identification
($\alpha < -0.2$), and that a large fraction of their detections would not
be classified
as detections in this catalog as their flux measurements do not exceed
$4\sigma$.
Given these limitations, the fact that we have an approximate $50\%$
overlap at the
other bands is in relative agreement with the range of co-detections of
\citet{pan02}.

\subsection{Nuclear Emission}

\begin{figure}
\begin{center}
\includegraphics[width=6in]{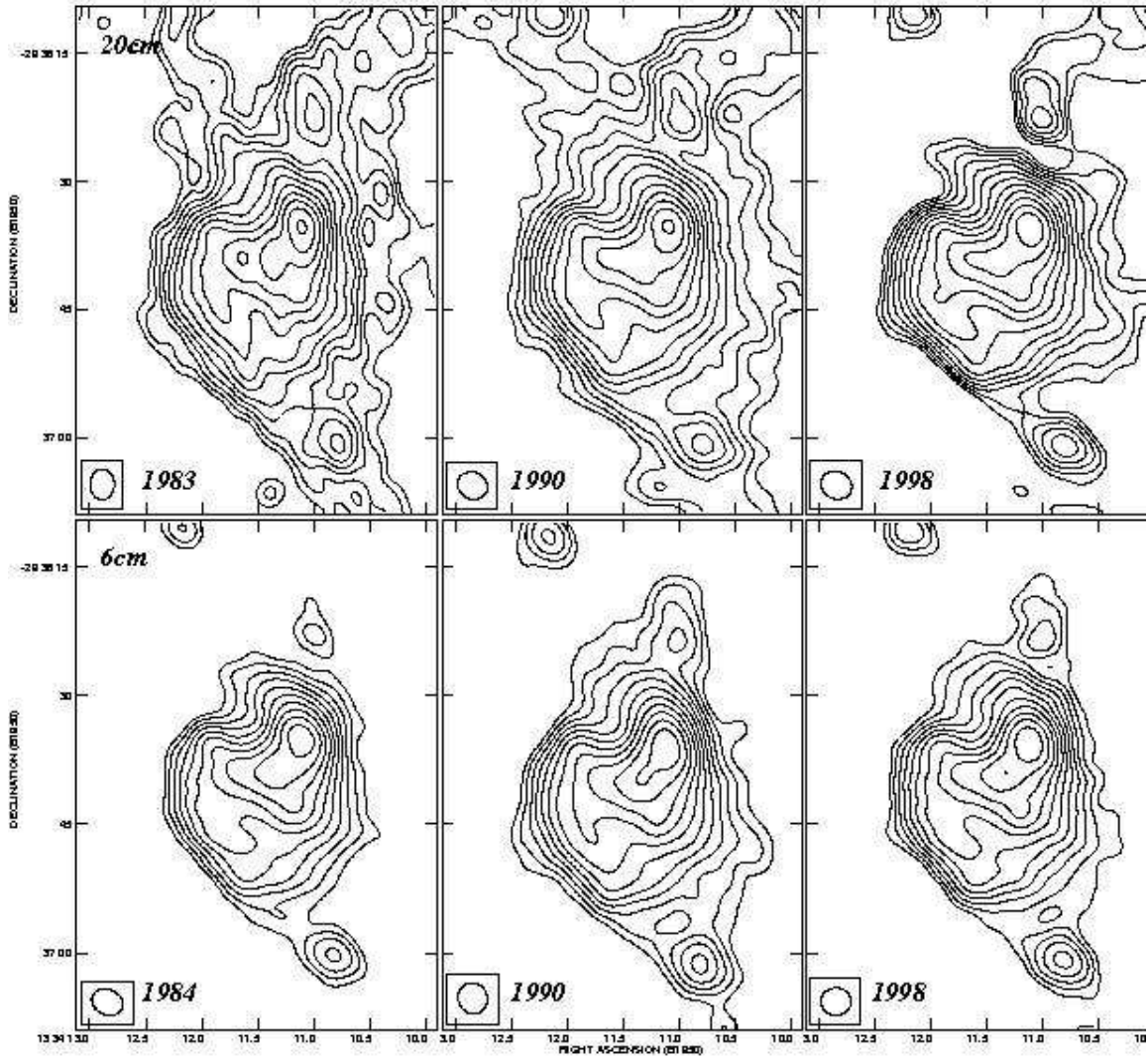}
\end{center}
\caption[{\sc Nuclear 20cm Contours}]{\label{nuctime}
Same as Figure \ref{57dtime}, but for the nuclear region. Contour levels are 0.40, 0.57, 0.8, 1.13, 1.60, 2.26,
3.20, 4.53, 6.40, 9.05, 12.80, 18.10 and 25.59 mJy/beam.}
\end{figure}

The nucleus of M83 is very complex.  The bright emission prevents the resolution of individual 
sources using the VLA configurations for the observation bands here.  \citet{crb} noted that
the optical/IR nucleus was not coincident with the radio nucleus.  We are able to confirm this 
with the new analysis of the data.  The peak of the radio nuclear emission is located at a distance
of $\sim 10^{\prime\prime}$ from the optical center. Contour plots of all maps are shown in Figure 
\ref{nuctime}.

\begin{figure}
\begin{center}
\includegraphics[height=7in]{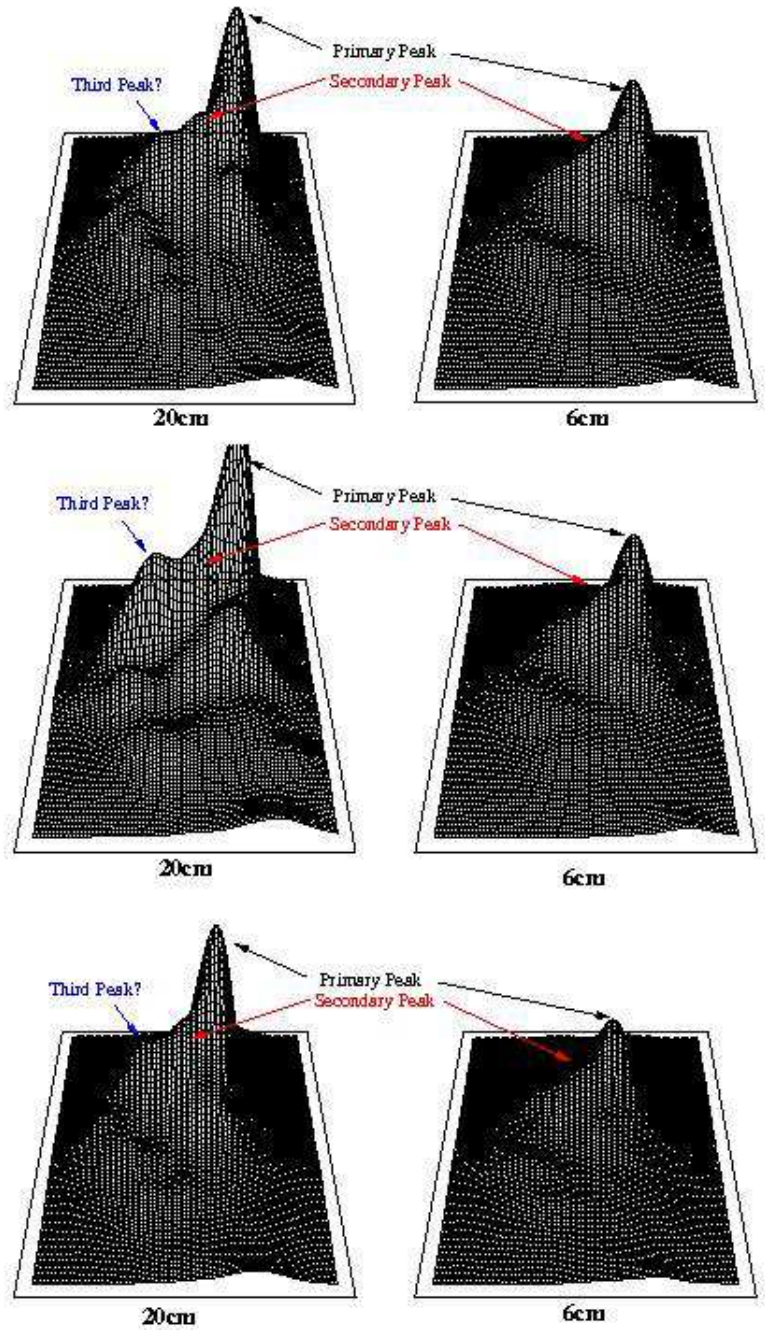}
\end{center}
\caption[{\sc Profile Plot of M83 Nucleus}]{
\label{m83profiles} A series of profile plots of the nuclear region.  Flux peaks are scaled to
the 1990 20cm observation to show the
differences between 20cm (left) and 6cm (right).  Time progresses from top to bottom.
One notable feature of these plots is the evidence for several lesser ``peaks'' of emission
in the 20cm images. 
}
\end{figure}

There does exist evidence for multiple emission sources within the nuclear complex.  From the
1983$-$1984 observations to those in 1992, there was an overall increase in the nuclear emission.
Figure \ref{m83profiles} shows a series of three dimensional emission profiles of the nuclear region.
The plots for the 20cm band indicate the presence of at least four possible sources.  The position
of the third brightest ``peak'' is consistent with the optical center.  \citet{soria02} report the 
presence of a hard X-ray source at this position.  X-ray spectral analysis \citep{soria02} indicate
that the source is consistent with a low accretion-rate black hole with mass $\sim 10^7 M_{\odot}$.
This compares favorably with the mass estimates of \citet{that00} ($1.6\times10^7M_{\odot}$) and
\citet{mast05} ($1.2\times10^7M_{\odot}$).
Although we cannot get accurate measurements of the nuclear peak flux levels in the radio maps, qualitative
inspection indicates that the emission is non-thermal in nature which would be consistent
with the X-ray results.

\begin{figure}
\begin{center}
\includegraphics[width=6in]{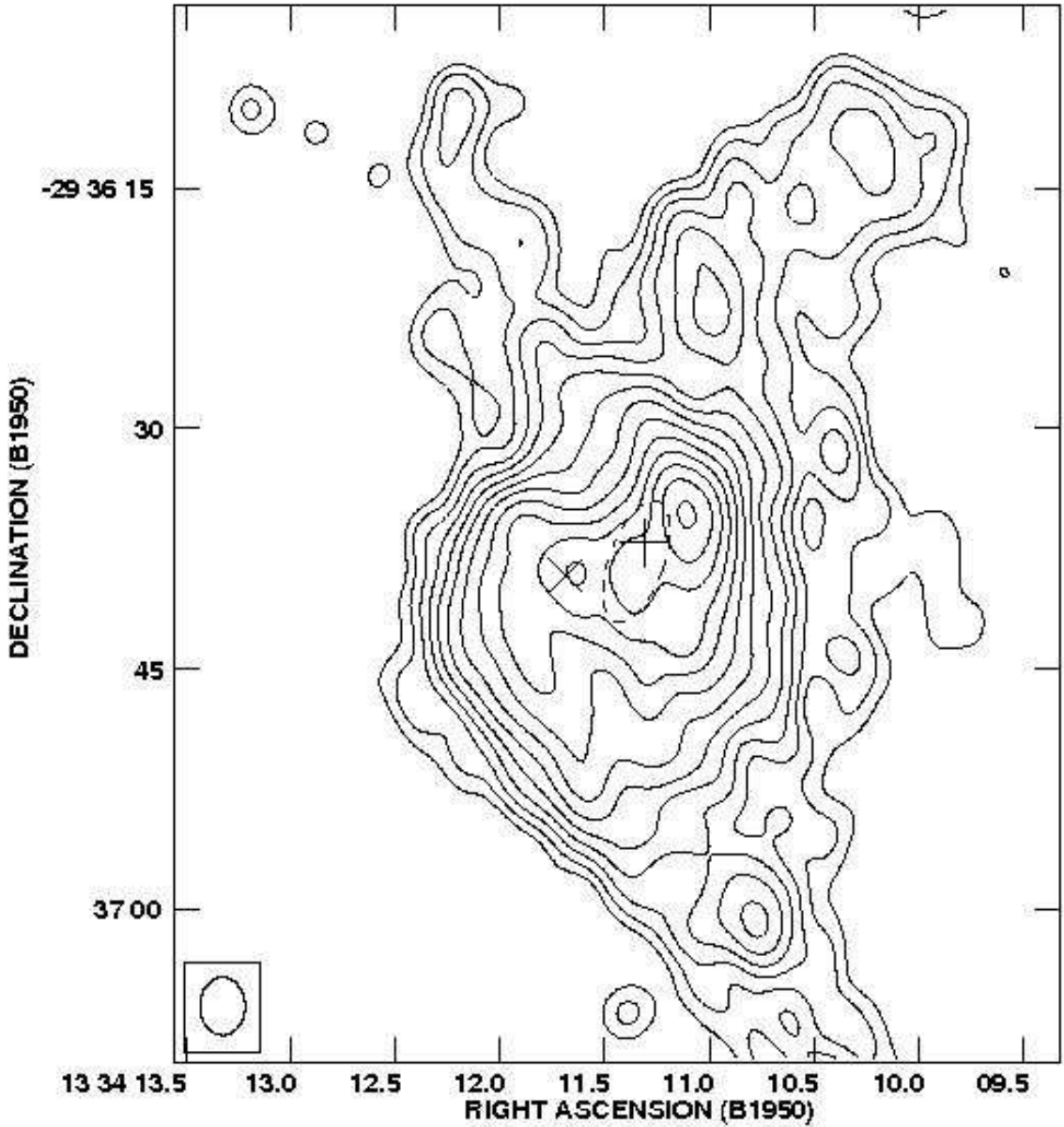}
\end{center}
\caption[{\sc Optical and Hidden Nucleus}]{\label{hidd}
Contours of 20cm emission from the 1983 observation.  Contour levels are same as Figure \ref{nuctime}.
The cross indicates the position of the optical nucleus.  The plus sign indicates the 
reported position of the hidden nuclear mass as determined by \citet{mast05}.  The
ellipse surrounding the plus indicates a probable region in which the hidden mass lies
due to another estimate by \citet{that00}.
}
\end{figure}

The second highest ``peak'' seems to be consistent with the obscured ``dark'' nucleus reported
by \citet{that00} and subsequently studied in \citet{sak04}.  In a later study, however, \citet{mast05}
reported a different position for this ``dark'' nucleus.  Figure \ref{hidd} shows positions
of the optical nucleus and the ``dark'' nucleus on the 1983 20cm contours. 
As illustrated in Figure \ref{hidd} 
nucleus appears closer to the peak of the radio emission of the galaxy.  The studies of this mass concentration
indicate that it is more massive \citep[e.g., ][]{mast05} than the the optical nucleus.  \citet{that00}
also note that the ``dark'' nucleus is located near 
the dynamical center of the galaxy.  
The positions of the nuclear mass concentrations are given in Table
\ref{npos}.  The positions for the ``dark'' nucleus are offset in
Declination by $\sim3.6^{\prime\prime}$, which corresponds to the deconvolution
size
for our observations. We would need the higher resolution of VLBI
observations in order to isolate the radio emission sources.

A VLBI observation using the Long Baseline Array of the 
Australia Telescope has been performed on the nuclear region of M83. The aim of these observations
is to resolve the individual sources in this region to get a better idea of the distribution and
to find multiple counterparts to the nuclear X-ray sources reported in \citet{soria03}. The results
of this study will be presented in a future paper (Maddox et al, in preparation).

\section{Conclusions} 

We have studied the long term  radio emission from compact sources in the galaxy M83. 
The observations included previously unpublished data from 1990 and 1998.  
We have presented the data over the fifteen years in a consistent manner.
Our observations have resulted in the detection of a number of objects
in M83:

\noindent$\bullet$
It was shown that SN 1957D has continued fading, consistent with an expanding shock through
a circumstellar material that is decreasing in density.  SN 1950B has apparently faded to the 
level of thermal {\sc Hii} regions that are near the position of the explosion. SN 1923A has
faded to near the limits of detection of these observations.  We continue to show no detection
of SN 1983N after its initial radio detection, consistent with it having been a Type Ib
supernova.

\noindent$\bullet$
About half of the radio sources are thermal {\sc Hii} regions.  A result that is not surprising
due to the high star formation activity of M83. The {\sc Hii} regions
tend to be very large, exhibiting high excitation parameters.  The largest
regions are not detected due to the high resolution of our
observations.

\noindent$\bullet$
It was found that ten sources were coincident with X-ray sources.  The continuum spectral indices
of these sources indicated that most were X-ray supernova remnants.  We confirm that one of the 
coincident sources (source 28) is the nucleus of a background radio galaxy with two
radio lobes (sources 27 and 29).  Three of the X-ray sources are coincident with
known optical supernova remnants.

\noindent$\bullet$
We note that the nuclear region of M83 has shown a slight increase in 20 cm emission in the
radio peak.  It is possible that there is an increase in accretion onto a supermassive black
hole, which would be consistent with X-ray results.  The 6 cm emission does not show this
increase in flux.
We confirmed that the reported optical/IR nuclear peak is not consistent with the radio
nucleus, though there is evidence for a radio emission region at the position of the optical
nucleus.  It was seen that the nuclear radio peak was near the position of a second ``dark''
nuclear mass concentration that corresponds to the dynamical nucleus of the galaxy.

Our
multiwavelength analyses have provided new information about the
nature of the many, detected compact sources and the nucleus in the
starburst galaxy M83.
Studies of the compact sources are providing new information about
late-term stellar evolution ({\it e.g.}, mass-loss rates),  emission
mechanisms, the transition of SNe into SNRs,
and the nature of XRBs. The combination of radio, X-ray and optical observations provide an 
excellent diagnostic for the classification of compact sources in this and other nearby 
galaxies.  This galaxy also appears to have
a complicated nuclear structure, including possibly
multiple supermassive black holes, and may have minor
variability.
In the future we will
continue to follow the long-term evolution of these sources in this
galaxy, as well as to compare the results here
with those in other nearby face-on
spiral galaxies.

We thank the referee, A. Pedlar,
for useful comments that have helped us to improve the paper.
This work employed extensive use of the NASA Extragalactic Database (NED).
This work was supported by NSF Grant AST-03-07279 (JJC). CJS is a
Cottrell Scholar of Research Corporation and work on this              
project has been supported by the NASA Wisconsin Space Grant Consortium.


\clearpage


\begin{deluxetable}{lcccccc}
\tabletypesize{\footnotesize}
\tablecaption{{\sc \label{obs} VLA Radio Observations of M83}}
\tablehead{
\colhead{}          &   \multicolumn{2}{c}{$1983-1984$ {\sc Observations}} & 
\multicolumn{2}{c}{1990 {\sc Observations}} & \multicolumn{2}{c}{1998 
{\sc Observations}} \\
\cline{2-7} 
\colhead{{\sc Parameter}} &  \colhead{20cm} & \colhead{6cm} & \colhead{20cm}
 & \colhead{6cm} & \colhead{20cm} & \colhead{6cm} 
}
\startdata
Frequency (GHz)\dotfill &  1.446  & 4.873   & 1.452 & 4.873 & 1.465  & 4.885\\
Observing dates\dotfill&  1983 Dec 15,17 & 1984 Mar 15   & 1990 Jun 30 & 1990 Oct 14 & 1998 Jun 13  & 1998 Oct 31\\
Observing time (hr)\dotfill&  3  & 6.5   & 7.5 & 6 & 11.3  & 11.6 \\
Configuration\dotfill&  BnA  & CnB   & BnA & CnB & BnA  & CnB\\
Primary beam HPBW\dotfill&  $30^\prime$  & $8^\prime$   &$30^\prime$  & $8^\prime$ & $30^\prime$  & $8^\prime$ \\
Clean Beam\dotfill&
$3.\!\!^{\prime\prime}50\times3.\!\!^{\prime\prime}50$ &
$3.\!\!^{\prime\prime}93\times2.\!\!^{\prime\prime}80$ &
$3.\!\!^{\prime\prime}65\times3.\!\!^{\prime\prime}65$ &
$3.\!\!^{\prime\prime}65\times3.\!\!^{\prime\prime}65$ &
$3.\!\!^{\prime\prime}65\times3.\!\!^{\prime\prime}65$ &
$3.\!\!^{\prime\prime}65\times3.\!\!^{\prime\prime}65$ \\
rms noise (mJy ${\mathrm{beam}}^{-1}$)\dotfill&
0.197 &
0.045 &
0.074 &
0.037 &
0.079 &
0.070 \\
\enddata
\end{deluxetable}

\clearpage

\begin{deluxetable}{lllcccccc}
\tablewidth{0pt}
\tabletypesize{\scriptsize}
\tablecolumns{9}
\tablecaption{
\label{flux}
Radio Positions and Peak Flux Densities of Sources in M83}
\tablehead{\colhead{}      & \multicolumn{2}{c}{\raisebox{-1.5ex}[0pt][0pt]{Position\tablenotemark{a}}} &
\multicolumn{2}{c}{1983-1984\tablenotemark{b}} & \multicolumn{2}{c}{1990} &
\multicolumn{2}{c}{1998} \\
\cline{4-9} \\
\cline{2-3}
\multicolumn{3}{c}{}    & \colhead{20 cm} & \colhead{6 cm}      &
\colhead{20 cm}         & \colhead{6 cm}  & \colhead{20 cm}     &
\colhead{6 cm}          \\
\colhead{Source}       & \colhead{R.A.(2000)}  & \colhead{Decl.(2000)} &
\colhead{(mJy)}         & \colhead{(mJy)} & \colhead{(mJy)}     &
\colhead{(mJy)}         & \colhead{(mJy)} & \colhead{(mJy)}
}

\startdata
1 & $13^{\mathrm{h}}36^{\mathrm{m}}50^{\mathrm{s}}\!\!.00$ & $-29^{\circ}52'43''\!\!.36$ & $\cdots$ & $0.41\pm0.08$ & $0.37\pm0.08$ & $0.35\pm0.04$ & $0.52\pm0.07$ & $0.34\pm0.06$ \\
2  & 13  36   50.83  & $-$29  51   59.56  &$\cdots$  & $\cdots$ & $0.38\pm0.08$ & $\cdots$ & $\cdots$ & $\cdots$ \\
3  & 13  36   50.86  & $-$29  52   38.54 & $0.76\pm0.20$ & $0.39\pm0.08$ & $0.38\pm0.08$ & $0.25\pm0.03$ & $0.40\pm0.09$ & $0.20\pm0.06$ \\
4  & 13  36   51.11  & $-$29  50   41.98 & $0.96\pm0.19$ & $\cdots$ & $0.55\pm0.08$ & $0.23\pm0.04$ & $0.58\pm0.07$ & $0.34\pm0.04$ \\
5\tablenotemark{c}  & 13  36   51.24  & $-$29  54   02.01  & $4.58\pm0.20$ & $1.30\pm0.12$ & $\cdots$ & $\cdots$ & $\cdots$ & $\cdots$ \\
6  & 13  36   51.55  & $-$29  53   00.61 & $\cdots$ & $\cdots$ & $0.43\pm0.08$ & $0.41\pm0.05$ & $0.54\pm0.10$ & $0.36\pm0.06$ \\
7  & 13  36   52.78  & $-$29  52   31.58 &$\cdots$  & $\cdots$ & $0.52\pm0.08$ & $0.25\pm0.04$ & $0.47\pm0.10$ & $0.29\pm0.05$ \\
8  & 13  36   52.77  & $-$29  51   10.39 & $\cdots$ & $0.45\pm0.07$ & $0.38\pm0.08$ & $0.49\pm0.04$ & $0.42\pm0.09$ & $0.30\pm0.05$ \\
9  & 13  36   52.83  & $-$29  51   37.96 & $\cdots$ & $\cdots$ & $0.62\pm0.08$ & $\cdots$ & $0.90\pm0.08$ & $\cdots$ \\
10  & 13  36   52.91  & $-$29  52   49.07 & $0.73\pm0.19$ & $0.39\pm0.08$ & $0.60\pm0.08$ & $0.28\pm0.05$ & $0.45\pm0.09$ & $0.28\pm0.05$ \\
11\tablenotemark{c}  & 13  36   52.92  & $-$29  51   56.50 & $0.70\pm0.17$ & $0.43\pm0.06$ & $0.54\pm0.08$ & $0.54\pm0.05$ & $0.53\pm0.07$ & $0.50\pm0.04$ \\
12  & 13  36   53.14  & $-$29  51   33.12 & $0.98\pm0.19$ & $1.03\pm0.06$ & $0.95\pm0.08$ & $1.16\pm0.05$ & $0.89\pm0.13$ & $1.17\pm0.05$ \\
13  & 13  36   53.27  & $-$29  52   57.44 & $0.47\pm0.19$ & $0.53\pm0.08$ & $0.45\pm0.08$ & $0.40\pm0.06$ & $0.60\pm0.08$ & $0.47\pm0.05$ \\
14  & 13  36   53.25  & $-$29  50   58.62 & $\cdots$ & $0.31\pm0.06$ & $0.42\pm0.08$ & $0.41\pm0.04$ & $0.20\pm0.08$ & $0.35\pm0.04$ \\
15  & 13  36   53.39  & $-$29  51   11.17 & $\cdots$ & $\cdots$ & $\cdots$ & $0.18\pm0.04$ & $\cdots$ & $0.19\pm0.05$ \\
16  & 13  36   54.20  & $-$29  50   42.22 & $\cdots$ & $0.45\pm0.06$ & $0.37\pm0.08$ & $0.30\pm0.05$ & $0.41\pm0.07$ & $0.33\pm0.04$ \\
17  & 13  36   54.40  & $-$29  53   05.17 & $0.43\pm0.21$ & $0.37\pm0.08$ & $0.36\pm0.08$ & $0.49\pm0.05$ & $0.52\pm0.11$ & $0.58\pm0.06$ \\
18  & 13  36   54.74  & $-$29  52   56.77 & $0.59\pm0.19$ & $0.47\pm0.08$ & $0.60\pm0.08$ & $0.33\pm0.05$ & $0.44\pm0.10$ & $0.29\pm0.06$ \\
19  & 13  36   54.90  & $-$29  53   10.29 & $\cdots$ & $0.58\pm0.08$ & $\cdots$ & $0.40\pm0.04$ & $0.36\pm0.10$ & $0.43\pm0.05$ \\
20  & 13  36   54.95  & $-$29  52   40.21 & $\cdots$ & $\cdots$ & $\cdots$ & $0.19\pm0.05$ & $0.28\pm0.09$ & $\cdots$ \\
21  & 13  36   55.41  & $-$29  52   56.08 & $\cdots$ & $\cdots$ & $0.26\pm0.08$ & $\cdots$ & $0.26\pm0.09$ & $0.21\pm0.03$ \\
22  & 13  36   55.54  & $-$29  53   03.26 & $\cdots$ & $\cdots$ & $0.22\pm0.08$ & $0.26\pm0.04$ & $0.33\pm0.09$ & $0.32\pm0.05$ \\
23  & 13  36   55.72  & $-$29  49   52.14 & $0.88\pm0.18$ & $0.51\pm0.06$ & $0.70\pm0.08$ & $0.43\pm0.04$ & $0.46\pm0.06$ & $0.53\pm0.05$ \\
24  & 13  36   56.13  & $-$29  52   54.99 & $\cdots$ & $\cdots$ & $0.45\pm0.08$ & $0.31\pm0.04$ & $0.42\pm0.07$ & $\cdots$ \\
25  & 13  36   56.32  & $-$29  49   34.10 & $\cdots$ & $\cdots$ & $\cdots$ & $0.13\pm0.03$ & $0.30\pm0.05$ & $\cdots$ \\
26  & 13  36   56.84  & $-$29  52   48.36 & $\cdots$ & $\cdots$ & $\cdots$ & $0.27\pm0.04$ & $0.35\pm0.08$ & $0.25\pm0.04$ \\
27  & 13  36   56.91  & $-$29  50   43.28 & $3.39\pm0.18$ & $0.86\pm0.05$ & $3.23\pm0.08$ & $1.09\pm0.03$ & $3.38\pm0.07$ & $1.02\pm0.04$ \\
28  & 13  36   58.34  & $-$29  51   04.58 & $0.74\pm0.18$ & $1.31\pm0.06$ & $1.06\pm0.08$ & $1.55\pm0.04$ & $0.82\pm0.07$ & $0.87\pm0.05$ \\
29  & 13  36   59.00  & $-$29  51   16.04 & $1.39\pm0.21$ & $0.37\pm0.05$ & $1.45\pm0.07$ & $0.60\pm0.04$ & $1.38\pm0.07$ & $0.49\pm0.04$ \\
30  & 13  36   59.54  & $-$29  51   26.32 & $\cdots$ & $\cdots$ & $0.73\pm0.07$ & $\cdots$ & $1.26\pm0.09$ & $\cdots$ \\
31  & 13  36   59.98  & $-$29  52   16.65 & $2.29\pm0.19$ & $1.35\pm0.06$ & $2.19\pm0.08$ & $1.45\pm0.05$ & $2.11\pm0.09$ & $1.56\pm0.05$ \\
32  & 13  37   00.17  & $-$29  51   40.02 & $1.75\pm0.21$ & $0.42\pm0.06$ & $2.16\pm0.08$ & $0.57\pm0.03$ & $2.85\pm0.12$ & $0.38\pm0.08$ \\
33  & 13  37   00.54  & $-$29  54   18.50 & $\cdots$ & $\cdots$ & $0.20\pm0.08$ & $0.27\pm0.05$ & $\cdots$ & $\cdots$ \\
34  & 13  37   01.34  & $-$29  51   26.51 & $1.34\pm0.22$ & $0.95\pm0.05$ & $0.92\pm0.08$ & $0.87\pm0.04$ & $1.14\pm0.09$ & $0.85\pm0.04$ \\
35  & 13  37   02.36  & $-$29  51   25.85 & $0.94\pm0.21$ & $\cdots$ & $0.77\pm0.08$ & $0.48\pm0.04$ & $0.66\pm0.08$ & $0.37\pm0.05$ \\
36  & 13  37   03.24  & $-$29  52   26.56 & $\cdots$ & $1.13\pm0.06$ & $0.50\pm0.08$ & $0.63\pm0.04$ & $0.68\pm0.07$ & $0.32\pm0.04$ \\
37  & 13  37   03.28  & $-$29  51   13.7  & $\cdots$ & $\cdots$ & $0.24\pm0.08$ & $\cdots$ & $\cdots$ & $0.22\pm0.05$ \\
38  & 13  37   03.39  & $-$29  54   02.04 & $1.01\pm0.23$ & $1.07\pm0.07$ & $0.66\pm0.08$ & $1.04\pm0.05$ & $0.80\pm0.06$ & $1.00\pm0.05$ \\
39\tablenotemark{c}  & 13  37   03.53  & $-$29  49   40.56 & $2.72\pm0.21$ & $2.18\pm0.06$ & $1.88\pm0.08$ & $1.57\pm0.04$ & $0.82\pm0.09$ & $0.60\pm0.05$ \\
40  & 13  37   04.73  & $-$29  50   57.59 & $\cdots$ & $0.32\pm0.06$ & $0.27\pm0.08$ & $0.36\pm0.04$ & $0.67\pm0.08$ & $0.41\pm0.05$ \\
41  & 13  37   06.61  & $-$29  53   32.32 & $\cdots$ & $0.75\pm0.09$ & $0.59\pm0.08$ & $0.42\pm0.05$ & $0.56\pm0.08$ & $0.31\pm0.05$ \\
42  & 13  37   06.91  & $-$29  49   36.03 & $1.30\pm0.23$ & $1.42\pm0.05$ & $1.51\pm0.08$ & $1.53\pm0.04$ & $1.54\pm0.07$ &$1.50\pm0.05$ \\
43  & 13  37   07.37  & $-$29  51   06.38 & $\cdots$ & $0.32\pm0.06$ & $0.27\pm0.08$ & $0.43\pm0.04$ & $0.38\pm0.06$ & $0.34\pm0.05$ \\
44  & 13  37   07.40  & $-$29  52   06.91 & $\cdots$ & $\cdots$ & $0.23\pm0.08$ & $0.20\pm0.04$ & $\cdots$ & $\cdots$ \\
45  & 13  37   07.66  & $-$29  51   13.28 & $0.60\pm0.21$ & $0.31\pm0.06$ & $0.55\pm0.07$ & $0.47\pm0.04$ & $0.49\pm0.07$ & $0.40\pm0.06$ \\
46  & 13  37   07.74  & $-$29  53   13.99 & $\cdots$ & $\cdots$ & $0.26\pm0.08$ & $0.22\pm0.04$ & $\cdots$ & $0.23\pm0.04$ \\
47  & 13  37   07.82  & $-$29  52   41.69 & $\cdots$ & $\cdots$ & $0.28\pm0.08$ & $0.26\pm0.05$ & $0.23\pm0.06$ &$\cdots$ \\
48  & 13  37   07.89  & $-$29  51   17.78 & $0.66\pm0.22$ & $0.65\pm0.07$ & $0.98\pm0.07$ & $0.53\pm0.04$ & $0.65\pm0.07$ & $0.62\pm0.05$ \\
49  & 13  37   08.09  & $-$29  52   55.78 & $\cdots$ & $\cdots$ & $0.34\pm0.09$ & $0.23\pm0.05$ & $0.34\pm0.07$ & $0.29\pm0.05$ \\
50  & 13  37   08.31  & $-$29  52   11.54 & $0.68\pm0.22$ & $0.35\pm0.08$ & $0.46\pm0.09$ & $0.41\pm0.04$ & $0.40\pm0.08$ & $0.38\pm0.05$ \\
51  & 13  37   08.69  & $-$29  51   31.04 & $\cdots$ & $0.39\pm0.06$ & $0.26\pm0.08$ & $0.31\pm0.04$ & $0.48\pm0.08$ & $0.23\pm0.05$ \\
52  & 13  37   09.19  & $-$29  51   33.31 & $0.86\pm0.22$ & $0.38\pm0.07$ & $0.48\pm0.07$ & $0.33\pm0.04$ & $0.70\pm0.07$ & $0.35\pm0.04$ \\
53\tablenotemark{c}  & 13  37   09.22  & $-$29  51   00.51 & $\cdots$ & $\cdots$ & $\cdots$ & $0.28\pm0.04$ & $0.22\pm0.06$ & $0.19\pm0.05$ \\
54  & 13  37   10.01  & $-$29  51   28.24 & $\cdots$ & $0.45\pm0.06$ & $\cdots$ & $0.20\pm0.04$ & $0.29\pm0.09$ & $0.25\pm0.06$ \\
55  & 13  37   11.44  & $-$29  49   52.31 & $\cdots$ & $\cdots$ & $\cdots$ & $0.23\pm0.05$ & $0.21\pm0.06$ &  $0.18\pm0.06$\\
\enddata
\tablenotetext{a}{From 1998 6 cm observations.}
\tablenotetext{b}{Flux measurements using original maps discussed in \citet{cb85}.  Included here
for completeness.}
\tablenotetext{c}{In order of appearance:  SN 1983N, SN 1950B, SN 1957D, and SN 1923A.}
\end{deluxetable}

\clearpage

\begin{deluxetable}{lr@{$\pm$}rr@{$\pm$}r}
\tablewidth{0pt}
\tablecaption{
{\sc \label{spix} Spectral Indices of Sources in M83}}
\tablehead{\colhead{}      & \multicolumn{4}{c}{{\sc Spectral Index}} \\
\cline{2-5}
\colhead{{\sc Source}} &  \multicolumn{2}{c}{1990} &
\multicolumn{2}{c}{1998} }
\startdata 
 1 &  -0.05 &   0.24 &  -0.36 &   0.22 \\ 
 2 &   \multicolumn{2}{c}{$<-1.01$}     &   \multicolumn{2}{c}{$\cdots$} \\
 3 &  -0.35 &   0.24 &  -0.58 &   0.38 \\
 4 &  -0.73 &   0.23 &  -0.45 &   0.17 \\
 5 &   \multicolumn{2}{c}{$\cdots$}  &   \multicolumn{2}{c}{$\cdots$} \\
 6 &  -0.04 &   0.22 &  -0.34 &   0.25 \\
 7 &  -0.61 &   0.22 &  -0.41 &   0.27 \\
 8 &   0.21 &   0.23 &  -0.28 &   0.27 \\
 9 &   \multicolumn{2}{c}{$<-1.42$ }    & \multicolumn{2}{c}{$<-1.20$ }      \\
10 &  -0.64 &   0.22 &  -0.40 &   0.27 \\
11 &   0.00 &   0.17 &  -0.05 &   0.15 \\
12 &   0.17 &   0.09 &   0.23 &   0.15 \\
13 &  -0.10 &   0.23 &  -0.21 &   0.17 \\
14 &  -0.02 &   0.21 &   0.47 &   0.42 \\
15 &   \multicolumn{2}{c}{$>-0.17$ }    &   \multicolumn{2}{c}{$>-0.18$ }    \\
16 &  -0.18 &   0.27 &  -0.18 &   0.21 \\
17 &   0.26 &   0.24 &   0.09 &   0.24 \\
18 &  -0.50 &   0.20 &  -0.35 &   0.31 \\
19 &   \multicolumn{2}{c}{$>0.49$ }     &   0.15 &   0.30 \\
20 &  \multicolumn{2}{c}{ $>-0.13$ }    &   \multicolumn{2}{c}{$<-0.23$ }    \\
21 &   \multicolumn{2}{c}{$<-0.70$ }    &  -0.18 &   0.37 \\
22 &   0.14 &   0.39 &  -0.03 &   0.31 \\
23 &  -0.41 &   0.15 &   0.12 &   0.16 \\
24 &  -0.31 &   0.22 &  \multicolumn{2}{c}{$<-0.57$ }     \\
25 &   \multicolumn{2}{c}{$>-0.44$ }    &  \multicolumn{2}{c}{$<-0.29$ }     \\
26 &   \multicolumn{2}{c}{$>0.16$ }     &  -0.28 &   0.28 \\
27 &  -0.91 &   0.04 &  -1.01 &   0.04 \\
28 &   0.32 &   0.08 &   0.05 &   0.10 \\
29 &  -0.74 &   0.08 &  -0.87 &   0.10 \\
30 &  \multicolumn{2}{c}{$<-1.55$ }     &  \multicolumn{2}{c}{$<-1.48$ }     \\
31 &  -0.35 &   0.05 &  -0.25 &   0.05 \\
32 &  -1.12 &   0.06 &  -1.70 &   0.21 \\
33 &   0.25 &   0.44 &   \multicolumn{2}{c}{$\cdots$}  \\
34 &  -0.05 &   0.10 &  -0.25 &   0.09 \\
35 &  -0.40 &   0.13 &  -0.49 &   0.18 \\
36 &   0.19 &   0.17 &  -0.63 &   0.16 \\
37 &  \multicolumn{2}{c}{$<-0.64$ }     &  \multicolumn{2}{c}{$>-0.06$ }     \\
38 &   0.38 &   0.13 &   0.19 &   0.09 \\
39 &  -0.15 &   0.05 &  -0.26 &   0.14 \\
40 &   0.24 &   0.32 &  -0.41 &   0.17 \\
41 &  -0.28 &   0.18 &  -0.50 &   0.22 \\
42 &   0.01 &   0.06 &  -0.02 &   0.06 \\
43 &   0.39 &   0.31 &  -0.09 &   0.22 \\
44 &  -0.12 &   0.40 &   \multicolumn{2}{c}{$\cdots$}  \\
45 &  -0.13 &   0.15 &  -0.17 &   0.21 \\
46 &  -0.14 &   0.36 &   \multicolumn{2}{c}{$>-0.02$ }    \\
47 &  -0.06 &   0.34 &   \multicolumn{2}{c}{$<-0.08$ }    \\
48 &  -0.52 &   0.10 &  -0.04 &   0.13 \\
49 &  -0.33 &   0.34 &  -0.13 &   0.27 \\
50 &  -0.10 &   0.22 &  -0.04 &   0.24 \\
51 &   0.15 &   0.33 &  -0.62 &   0.27 \\
52 &  -0.31 &   0.19 &  -0.58 &   0.15 \\
53 &   \multicolumn{2}{c}{$>0.19$  }    &  -0.12 &   0.38 \\
54 &   \multicolumn{2}{c}{$>-0.09$ }    &  -0.12 &   0.39 \\
55 &   \multicolumn{2}{c}{$>0.03$  }    &  -0.13 &   0.44 \\
\enddata
\end{deluxetable}

\clearpage

\begin{table}
\caption[Optical Supernova Remnants]{\label{snrcount}
Optical Supernova Remnants with Radio Counterparts}
\begin{center}
\begin{tabular}{lcc}
\hline\hline
Source & Radio Spectral Index & SNR  \\
\hline
19\dotfill  & $+0.15\pm0.30$ & BL18 \\
22\dotfill  & $-0.18\pm0.37$ & BL24 \\
35\dotfill  & $-0.49\pm0.18$ & BL41 \\
39\dotfill  & $-0.26\pm0.14$ & BL50 \\
\hline
\end{tabular}
\end{center}
\end{table}

\clearpage

\begin{table}
\caption[Optical {\scshape Hii} Regions with Radio Counterparts]{\label{hiicount}
Radio Sources with Associated {\scshape Hii} Regions}
\begin{center}
\begin{tabular}{lclc}
\hline
Source & Radio Spectral Index & HII Region & Excitation Parameter\tablenotemark{a} \\
\hline\hline
2 & $<-1.02$ & RK230 & 241 \\
3 & $-0.58\pm0.38$ & RK233 & 355 \\
4 & $-0.45\pm0.17$ & RK223 & 444 \\
6 & $+0.34\pm0.25$ & RK292/295 & 228/287 \\
8 & $-0.28\pm0.27$ & RK211 & 393 \\
12 & $+0.23\pm0.15$ & RK209 & 321 \\
18\tablenotemark{b} & $-0.35\pm0.31$ & $\cdots$ & $\cdots$ \\
22 & $-0.03\pm0.31$ & RK185 & 369 \\
26 & $-0.28\pm0.28$ & RK181 & 201 \\
33 & $+0.25\pm0.44$ & RK144 & 300 \\
34 & $-0.25\pm0.09$ & RK137 & 473 \\
37 & $>-0.06$ & RK119 & 97 \\
40 & $-0.41\pm0.17$ & RK110 & 470 \\
42 & $-0.02\pm0.06$ & RK86 & 316 \\
43 & $-0.09\pm0.22$ & RK79 & 288 \\
45/48 & $-0.04\pm0.13$ & RK75 & 395 \\
46 & $-0.14\pm0.36$ & RK78 & 228 \\
49 & $-0.13\pm0.27$ & RK73 & 301 \\
50 & $-0.04\pm0.24$ & RK70 & 353 \\
53 & $-0.12\pm0.38$ & RK59 & 323 \\
55 & $-0.13\pm0.44$ & RK37 & 294 \\
\hline
\tablenotetext{a}{The excitation parameter $U$ is a measure of the total Lyman continuum flux and is a
useful index of the distribution of O stars.  The values are in units of pc cm$^{-2}$, as presented in \citet{rk83}.}
\tablenotetext{b}{This source was detected as an {\scshape Hii} in a survey by \citet{dev83}, listed as source 23.}
\end{tabular}
\end{center}
\end{table}

\clearpage

\begin{table}
\caption[X-Ray Counterparts to Radio Sources in M83]{
\label{xrcount}X-Ray Counterparts to Radio Sources in M83}
\begin{center}
\begin{tabular}{lcccc}
\hline\hline
Source &  Radio Spectral Index & CXOU & Source Type & Reference \\
\hline
3\dotfill & $-0.58\pm0.38$ & J133650.8-295240  &   {\scshape Hii} Region/SNR & 1\\
4\dotfill & $-0.45\pm0.17$ & J133651.1-295043  &   {\scshape Hii} Region/LMXB & 1\\
7\dotfill & $-0.41\pm0.27$ & J133652.8-295231  &   SNR & 1\\
19\dotfill & $+0.15\pm0.30$ & J133655.0-295239  &  SNR & 2\\
22\dotfill & $-0.02\pm0.31$ & J133655.6-295303  &  SNR & 2\\
24\dotfill & $-0.31\pm0.22$ & J133656.2-295255  &  $\cdots$\\
28\dotfill & $+0.05\pm0.10$ & J133658.3-295105  &  Background Galaxy & 3 \\
32\dotfill & $-1.70\pm0.21$ & J133700.0-295138  &  $\cdots$\\
35\dotfill & $-0.49\pm0.18$ & J133702.4-295126  &  SNR & 2\\
36\dotfill & $-0.63\pm0.16$ & J133703.3-295226  &  XRB\\
\hline
\end{tabular}
\end{center}

(1) \citet{colb04} \\
(2) \citet{blair04} \\
(3) \citet{poster1}

\end{table}

\clearpage

\begin{table}
\caption[{\sc Nuclear Mass Concentrations in M83}]{\label{npos} {\sc Reported Positions of Nuclear Mass Concentrations}}
\begin{center}
\begin{tabular}{lccc}
\hline\hline
                & RA (J2000) & Dec (J2000) & Reference \\
\hline
Optical Nucleus & 13 37 00.79 & $-$29 51 58.60 & \citet{crb} \\
Hidden Nucleus  & 13 37 00.54 & $-$29 51 53.62 & \citet{mast05} \\
                & 14 37 00.72 & $-$29 51 57.20 & \citet{that00} \\
\hline
\end{tabular}
\end{center}
\end{table}

\clearpage

\clearpage

\clearpage

\clearpage

\clearpage

\clearpage

\clearpage

\clearpage

\end{document}